\documentstyle [12pt]{article}
\def\o{\over}
\def\ra{\rightarrow}
\def\bar{\overline}
\def\r{\gamma}
\def\p{\phi}
\def\a{\alpha}
\def\b{\beta}

\def\G{{\rm GeV}}

\def\cba{{\bar c}_{L\alpha}}
\def\ba{b_{L\alpha}}
\def\bb{b_{L\beta}}
\def\sbb{{\bar s}_{L\beta}}
\def\sba{{\bar s}_{L\alpha}}
\begin{document}
\baselineskip=21pt
\setcounter{page}{1}
\thispagestyle{empty}
\begin{flushright}
\begin{tabular}{c c}
& {\normalsize   AUE-02-93}\\
& {\normalsize  EHU-03-93}\\
& {\normalsize  KGU-02-93}\\
& {\normalsize  July 1993}\\
& {\normalsize  revised January 1994}
\end{tabular}
\end{flushright}
\vspace{-0.3cm}
\centerline{\Large\bf Penguin-Diagram Induced $B \ra K_X \phi$ Decays}
\centerline{\Large\bf  in the Standard Model}
\centerline{\Large\bf  and}
\centerline{\Large\bf in the Two-Higgs-Doublet Model}
\vskip 0.3 cm
\centerline{{\bf Andrew J. DAVIES}$^{(a)}$,
{\bf Takemi HAYASHI}$^{(b)}$,
{\bf Masahisa MATSUDA}$^{(c)}$\footnote{E-mail:masa@auephyas.aichi-edu.ac.jp}}
\centerline{and}
\centerline{{\bf Morimitsu TANIMOTO}$^{(d)}$}
\vskip -0.3 cm
\centerline{$^{(a)}$ \it{Research School of Physical Sciences and
Engineering,}}
\centerline{\it Australian National University, Canberra ACT 0200, Australia}
\centerline{$^{(b)}$ \it{Kogakkan University, Ise, Mie 516, JAPAN}}
\centerline{$^{(c)}$ \it{Department of Physics and Astronomy, Aichi University
of Education}}
\centerline{\it Kariya, Aichi 448, JAPAN}
\centerline{$^{(d)}$ \it{Science Education Laboratory, Ehime University}}
\centerline{\it Matsuyama, Ehime 790, JAPAN}
\vskip 0.3cm
\centerline{\bf ABSTRACT}
\hspace*{0.6cm}
We systematically analyse the gluonic penguin-induced charmless decays
$B \ra K_X\phi$($K_X$ denotes the meson state
 $s\bar q \;(q=u\ {\rm or}\ d)$ ), in the standard model and
the two-Higgs-doublet model.
These processes, being induced at one-loop level, are of great
importance
in measuring the virtual top quark effect in the standard model, and also in
searching for the non-standard signals  in the low energy region.
It is shown that the QCD effect is also significant  in these processes,
as in the weak radiative processes $B \ra X_s \gamma$.
We also show that
the charged Higgs contribution can not provide  sizable enhancements
for the decays $B \ra K_X\phi$, in contrast to the decays $B \ra K_X\gamma$.
It is also found that processes such as $B \ra K_1(1400)\phi$
and $B \ra K(1460)\phi$ have
large branching fractions among $B \ra K_X\phi$ decays.
\newpage
\section{Introduction}
\hspace*{0.6cm}
Previously rare $B$ decays such as $b \ra s \gamma,
b \ra sl{\bar l}$,  $b \ra s\nu{\bar \nu}$,
$b{\bar q}\ra l{\bar l}$ and $b \ra sg$, have been
the subject of some interest in the literature.
The object of such studies has been to either allow confirmation
of the standard model(SM) or to find indications of additional
effects beyond the SM.
Taking such an approach in our previous paper
\cite{HAYASHI}(referred to as paper I), we analysed the inclusive decays
$B \ra X_s \gamma$ and exclusive decays $B \ra K_X\gamma$
(where $K_X$ denotes the meson state
$s\bar q \;({\bar q}={\bar u}\ {\rm or}\ {\bar d})$ ).
In this work we included the nonstandard physical effects due
to the charged
Higgs contribution in the  two-Higgs-doublet model(THDM). We
showed that sizable enhancements for these processes
are possible compared to the SM.
It is well known that, in these processes
QCD corrections play an important role
\cite{QCD}  in the quantitative
discussion, resulting in an enhancement of three to four times
in the branching ratio for $b \ra s \gamma$.

In the present paper, we discuss another interesting
class of processes: $B \ra K_X\phi$.
There is no previous work with full QCD corrections and including
higher $K$-meson resonances.
The characteristic feature of these processes is that they are mainly
induced through the gluonic penguin interaction via $b \ra s \ g^*$,
where $g^*$ denotes the virtually emitted gluon.
The non-leptonic $B$-decay processes to the final states including
$s+s+{\bar s}$ have previously been discussed by three of the present authors
and others, with the aim of clarifying the nonstandard physical
effects due to the charged Higgs contribution in the THDM
\cite{DAVIES}.
However, QCD corrections were not fully
included in these works, and the decays into
higher $K$-resonances with the $\phi$-meson were not included.

Now we want to make a more detailed analysis of the processes
$B \ra K_X\phi$.
First we summarize the QCD corrections to the effective Hamiltonian
for the process $b \ra s+s+{\bar s}$. For this process,
the QCD corrections induced by the dominant contribution of the
$t$-quark up to the one-loop level essentially require the additional
dimension-six local operators of Wilson's operator-product expansion.
These operators have been approximately neglected in the analyses of the other
rare $B$ decays
\cite{GRIN}.
Such consideration of more complete QCD corrections has already been
given by
Buchalla, Buras and Harlander
\cite{BUCHALLA}, who provided a detailed
renormalization group analysis to investigate
the ratio $\epsilon^{\prime}/\epsilon$ systematically by including the
contributions from the gluonic, photonic,
$Z^0$, and neutral Higgs penguin diagrams.
They obtained the Wilson coefficient functions
with full $O(\alpha_{QED})$ contributions in a compact and transparent form.

 The concrete and detailed description of our evaluation of the relevant
Wilson functions of the linearly independent dimension six
operators is given in section 2,
 although our evaluation of them is partly the same as that
made by
Buchalla, Buras and Harlander.
Note though that the estimation of the Wilson coefficient functions for
full operators in THDM
has not been previously given.
In section 3, we describe the calculation of the amplitudes and
the decay rates for $B \ra K_X\phi$.
We emphasize that our calculation of the amplitudes uses the
specific form factors
given by Isgur, Scora, Grinstein and Wise
\cite{ISGW},
as well as the factorization assumption
as  seen in this section.
Our results are compared with another model of form factors given by
Bauer, Stech and Wirbel
\cite{BSW}.
The effect of possible charged Higgs contributions is also studied.
Recently, new experimental results, in the form of a limit on and
a non-zero value of respectively, for $B(b \ra s\gamma)$
and $B(B \ra K^* \gamma)$, have been reported by
the CLEO experiment
\cite{CLEO}.
These results lead to new constraints on the values of the relevant parameters
occurring in expressions for the amplitudes of these decays.
Section 4 is devoted to brief comments on the resultant constraints.
In particular, we discuss the constraint on $\cot \beta$ versus $m_H$
in the THDM.
In section 5, our summary and conclusions are presented.

\section{QCD effects in the rare decay processes of $B$-meson}
The rare decay process of the  $b$-quark to the final state $s+s+{\bar s}$
is induced at the one-loop level mainly through $b \ra s+g^*$.

The fundamental four quark interaction mediated by the
process $b \ra s+g^*$, {\it i.e.} induced by the gluonic penguin diagram, is
described by the effective Hamiltonian
\begin{equation}
H_{penguin}=-{\alpha_s G_F \over 24\sqrt{2}\pi}V_{ts}^*V_{tb}{\bar s}(p)\{
\Gamma_\mu^{SM}+\Gamma_\mu^{2H}\}{\lambda^a \over 2}b(p+q)
{\bar q}(p_2)\gamma^\mu {\lambda^a \over2}q(p_1),
\end{equation}
where
\begin{equation}
\Gamma_\mu^{SM}=G_1(x_t)\gamma_\mu(1-\gamma_5)+3i{m_b \over q^2}G_2(x_t)\sigma
_{\mu\nu}q^\nu(1+\gamma_5)
\end{equation}
and
\begin{equation}
\Gamma_\mu^{2H}=F_1(y)\gamma_\mu(1-\gamma_5)+3i{m_b \over q^2}F_2(y)\sigma
_{\mu\nu}q^\nu(1+\gamma_5).
\end{equation}
The functions $G_i$ and $F_i$ are given by
\cite{INAMI}

\begin{eqnarray}
G_1(x_t)&=& {{x_t(1-x_t)(18-11x_t -
          x_t^2)-2(4-16x_t+9x_t^2)\ln x_t}\over
          (1-x_t)^4} \nonumber\\
&+&48\{\int_0^1dt\ t(1-t)\ln{m_c^2-q^2t(1-t)
\over m_W^2(1-t)+m_c^2t-q^2(1-t)}-{5 \over 36}\}\ , \nonumber\\
G_2(x_t)&=&x_t{{(1-x_t)(2+5x_t-x_t^2)+6x_t\ln x_t}\over (1-x_t)^4}\ ,
 \\
F_1(y)&=&y\cot^2\beta{{(1-y)(16-29y+7y^2)+6(2-3y)\ln y}\over
               {3(1-y)^4}}\ , \nonumber\\
F_2(y)&=&{1 \over 3}\cot^2\beta\ G_2(y)+2y{(1-y)(-3+y)-2\ln y \over (1-y)^3}\ ,
\nonumber
\end{eqnarray}
where $x_t=m_t^2/m_W^2, y=m_H^2/m_t^2$ and
$\cot\beta=v_d/v_u$ is the usual notation for the
 ratio of the vacuum expectation
values of the neutral sectors of the two-Higgs-doublets.
Here, in  the first equation for $G_1(x_t)$,
 we assume $V_{tb}=V_{cs},V_{ts}=V_{cb}$ and $V_{ub}=0$ .
Note that we have chosen a specific variant of the 2HDM, {\it i.e.} of
the form of the coupling of the Higgs
doublets to the fermions. There are four variants of the 2HDM,
distinguished by different schemes
for the Yukawa couplings to the fermion sector. The reader
will find a concise summary of the possible models in reference \cite{BHP}.
For what follows we will take model II of that work, in which one
doublet gives mass to the up-type quarks, and the other to the down-type
quarks and charged leptons. As we will discuss, there are considerable
uncertainties in the calculations that follow, and
as a consequence, we will content ourselves to
investigate the possibility of distinguishing between SM results and
those with
contributions arising from charged scalars.
Distinguishing further between competing models will
in general require considerable
refinements in the theoretical calculations.
We also note that, in the second term of the {\it r.h.s.},
we should include the charm quark contribution
due to the soft GIM cancellation. This term
reduces to the familiar result $8\ln m_c^2/m_W^2$ in the limit of
the squared momentum of virtual gluon vanishing; $q^2=0$.
Note though that  this limit is not realistic for the virtual gluonic penguin
induced processes, since the gluon is off mass-shell.
In the analysis to follow we take the appropriate nonzero value of $q^2$.

The following four-quark operators
and the magnetic-transition-type operators are relevant for the processes
under consideration. We write
the Hamiltonian
\begin{equation}
H_{eff}={4G_F \over \sqrt{2}}V_{tb}V^*_{ts}\sum_{i=1}^8 C_i(\mu)O_i(\mu)
\ ,
\end{equation}
where $\mu$ in the parentheses denotes the energy scale
at which the operators are
relevant for the decays.
The operators $O_i$ are defined as follows:
\begin{eqnarray}
           O_1 &=&({\bar c}_{L \alpha}\gamma^\mu \bb)(\sbb \gamma_\mu
c_{L\alpha})\ , \nonumber\\
           O_2 &=&(\cba\gamma^\mu \ba)(\sbb \gamma_\mu c_{L\beta}) \
,\nonumber\\
           O_3 &=&(\sba\gamma^\mu \ba)(\sum_{\rm 5\; quarks}{\bar q}_{L\beta}
            \gamma_\mu q_{L\beta}) \ ,\nonumber\\
           O_4 &=&(\sba\gamma^\mu \bb)(\sum_{\rm 5\; quarks}{\bar q}_{L\beta}
            \gamma_\mu q_{L\alpha}) \ ,\\
           O_5 &=&(\sba\gamma^\mu \ba)(\sum_{\rm 5\; quarks}{\bar q}_{R\beta}
            \gamma_\mu q_{R\beta}) \ ,\nonumber\\
           O_6 &=&(\sba\gamma^\mu \bb)(\sum_{\rm 5\; quarks}{\bar q}_{R\beta}
            \gamma_\mu q_{R\alpha}) \ ,\nonumber\\
           O_7 &=&-i{e \over {8\pi^2}}m_b\sba \sigma^{\mu\nu}b_{R\alpha}q_\mu
           \epsilon_\nu \ ,\nonumber\\
           O_8 &=&-i{g_c \over {8\pi^2}}m_b\sba
\sigma^{\mu\nu}T^a_{\alpha\beta}
           b_{R\beta}q_\mu \epsilon^a_\nu \ . \nonumber
\end{eqnarray}

Coefficients relevant to our processes are defined
at the energy scale $m_W$ to be
\begin{eqnarray}
C_1(m_W)&=&0 \ ,\nonumber\\
C_2(m_W)&=&1 \ ,\nonumber\\
C_3(m_W)&=&-{1 \over 3}C_4(m_W)=C_5(m_W)=-{1 \over 3}C_6(m_W) \nonumber\\
        &=&-{\alpha_s(m_W) \o 288\pi}(G_1(x_t)+F_1(y)) \ ,\\
C_7(m_W)&=&C_7^{SM}(x_t)+C_7^{2H}(y) \ ,\nonumber\\
C_8(m_W)&=&-{1 \o 8}(G_2(x_t)+F_2(y)). \nonumber
\end{eqnarray}
Here  $C_7^{SM}(x_t)$ and $C_7^{2H}(y)$ are
\begin{equation}
C_7^{SM}(x_t)={x_t \over {24(1-x_t)^3}}[8x_t^2+5x_t-7+{6x_t(3x_t-2)
\over (1-x_t)}\ln x_t]
\end{equation}
and
\begin{eqnarray}
C_7^{2H}(y) &=&{y \over 4(1-y)^3}[{5y^2-8y+3 \over 3}-{2 \over 3}
(3y-2)\ln y] \nonumber\\
&  &-\cot^2\beta{y \over 4(1-y)^4}[{8y^3-3y^2-12y+7 \over 18}+{2 \over
3}y(1-{3\ \over 2}y)\ln y] \ ,
\end{eqnarray}
respectively.
In Eqs.(7), we neglect the contributions from $Z^0$- and photonic-penguin
interactions
\cite{COM} since these terms are very small numerically due to $\alpha_s \ll
\alpha$ and do not affect the present analyses.
Also the second term, {\it i.e.} the charm contribution, in $G_1(x_t)$
should be taken into account
\cite{FLEI}
at the scale $\mu=m_b^2$ by setting $q^2$
$\simeq m_b^2/2$
\cite{q2}.

 Numerically, in the energy region between $m_W$ and $m_b$, we put the flavour
number $f=5$
in the renormalization group equation
\cite{MASA}.
We evolve the coefficients $C_i(\mu)$ starting from the scale $m_W$
as given in Eqs.(7) to
the scale $\mu=m_b=4.58$GeV, and then
we obtain the following numerical coefficients:
\begin{eqnarray}
C_1(m_b)&=&-0.240 \ ,\quad C_2(m_b)=1.103 \ ,\nonumber\\
C_3(m_b)&=&0.011+1.125C_3(m_W)-0.121C_4(m_W) \ ,\nonumber\\
C_4(m_b)&=&-0.025-0.291C_3(m_W)+0.824C_4(m_W) \ ,\nonumber\\
C_5(m_b)&=&0.007+0.944C_3(m_W)+0.083C_4(m_W) \ ,\\
C_6(m_b)&=&-0.030+0.229C_3(m_W)+1.465C_4(m_W) \ ,\nonumber\\
C_7(m_b)&=&-0.199+0.629C_3(m_W)+0.931C_4(m_W)+0.675C_7(m_W) \nonumber\\
        &+&0.091C_8(m_W) \ ,\nonumber\\
C_8(m_b)&=&-0.096-0.598C_3(m_W)+1.029C_4(m_W)+0.709C_8(m_W). \nonumber
\end{eqnarray}
In Table 1 we summarize the numerical values of the above coefficients
in the THDM for two values of the parameters $(m_H,\cot\beta)$ [denoted
in table by 2HDM(1) and 2HDM(2)],
as well as
the SM results for typical values of the SM parameters $V_{ij}$ and $m_t$.
In order to calculate the enhancement due to QCD corrections,
we have compared the coefficients in both cases with and without these
corrections.
Note that most of the previous predictions
for $B \ra K\phi$ and $K^* \phi$ decays have been given without the QCD
correction.
\vskip 0.5cm
\begin{center}
 \unitlength=0.8cm
 \begin{picture}(2.5,2.5)
  \thicklines
  \put(0,0){\framebox(3,1){\bf Table 1}}
 \end{picture}
\end{center}
\vskip 0.5cm
\noindent
In Table 1, the heading ``Without'' QCD means that
we include no RG evolution of the coefficients and fix their values
at the scale $m_W$, as given in  Eq.(7), with $q^2\simeq m_b^2/2$.
We see that the values of $C_i(m_b)$ with the QCD correction are
larger than those without by a factor of 1.5 $\sim$ 2.
The values $C_1(m_b) \sim C_6(m_b)$ are almost identical in the
SM, the THDM(1) and the THDM(2).
It is noteworthy that the differences between the three models
exist only for the coefficients $C_7(m_b)$ and $C_8(m_b)$.
As mentioned later in section 4, the value of $\cot \beta$ is strongly
 limited by the recent CLEO experimental search
 \cite{CLEO}
 for the $B \ra X_s \gamma$ decay.
As a result, here we use the values of $m_H$
and $\cot \beta$ which satisfy the present
 experimental restrictions.
In the next section we apply the above analysis on QCD corrections
to the calculation of  the decay amplitudes.
Note that no clear differences
among the SM and the THDMs for the coefficients $C_1(m_b) \sim C_6(m_b)$
are seen in Table 1 even using the QCD corrected coefficients.
So we expect that the decay $B \ra X_s \gamma$ will be the best process to
search for an effect beyond the SM.
However, it is still important to analyse the gluonic penguin effect
through the exclusive processes induced
by this interaction. This allows us to check the one-loop
 effects based on the SM, or to study the signals from beyond SM
physics. Such an analysis
is given in section 3.
\section{Exclusive decays $B \ra K_X\phi$ induced by the gluonic penguin
interaction}
\hspace*{0.6cm}
Of the penguin induced charmless $B$ meson decays,
 the exclusive non-leptonic rare  decays
  $B \ra K_X \phi$ are of special interest,
since these processes are the typical of those caused by the gluonic penguin
interaction.
Analysis of these exclusive decays is also helpful for the future
experimental study at $B$-factories.
In this section, we calculate these decay rates by including QCD corrections
described in section 2.
The contribution from new physics is also studied.
Since the form factors play an important role in estimating the branching
ratios precisely, we also study the form factor dependence of our results.
\par
  In the $s\bar q $ system,
  a rich spectrum of states has been observed.
The resonance state $K_X$ is specified by the quantum numbers
$n$, $L$, $s$ and $J$, which
denote the radial excitation quantum number, the orbital angular momentum,
the sum of the spins of the two quarks and the
total spin of the meson, respectively.
We investigate the following eight states with the notation $ n ^{2S+1}L_J $:
 $1 ^{1}S_0$,
$ 1 ^3S_1$, $1 ^3P_2$, $1 ^3 P_1$,  $1 ^1P_1$ , $1 ^3P_0$,
   $2 ^1S_0$ and  $2 ^3 S_1$.
Physical mesons corresponding to each state are
\cite{DATA}:
$$  1 ^{1}S_1:\ K\ ,\quad 1 ^{3}S_1:\ K^*(892)\ ,
\quad 1 ^3 P_2:\  K_2^*(1430)\ ,     $$
\begin{equation}  1 ^3 P_1\ ,   1 ^1P_1:\ K_1(1270)\ , \ K_1(1400)\ ,
\end{equation}
$$ 1 ^3P_0:\  K_0(1430)\ , \quad  2 ^1S_0:\  K(1460)\ ,
\quad 2 ^3 S_1:\ K^*(1410)\ . $$
The spin 1 mesons $K_1(1270)$ and $K_1(1400)$ are nearly $45^\circ$ mixed
 states of  $1 ^3 P_1$ and  $1 ^1P_1$.
\par
We proceed by showing
 the formulation to calculate the decay matrix elements.
The operators $O_3$, $O_4$, $O_5$, $O_6$ and $O_8'$
 are relevant for the decays $B \ra K_X\p$.
Here the operator $O_8'$ is derived
 from the interaction in which the virtual gluon in the magnetic operator $O_8$
 couples with $s+{\bar s}$, to give
\begin{equation}
           O_8' =-im_b{\alpha_s \over 2\pi}{1 \over q^2}
           \sba \sigma^{\mu\nu}T^a_{\alpha\beta}
           b_{R\beta}q_\mu {\bar s}_\beta \gamma_\nu
           T^a_{\beta\alpha}s^\alpha \ .
\end{equation}
By the use of these operators, the decay amplitude may be written as
\begin{equation} \langle K_X\p \mid H_{eff}\mid B\rangle
  = {4 G_F\over \sqrt{2}} V_{tb} V^*_{ts} \sum_{3,4,5,6,8'}C_i(\mu)
 \langle K_X\p \mid O_i(\mu) \mid B\rangle \ .
\end{equation}
\noindent
In the following analysis we take the value of the coefficient $C_8'(\mu)$
as equal to $C_8(\mu)$.
Although $C_8(\mu)$ does not include full the QCD correction of $C_8'(\mu)$
 in the leading {\it log} approximation, this replacement does not seriously
affect the results numerically, as the $O_8'$ term is the next to leading one,
compared to the operators $O_3, O_4, O_5$ and $O_6$.

We also use the factorization approximation
in order to estimate
 the hadronic matrix element.
The factorization assumption works successfully
 in $D$ meson  and $B$ meson decays
\cite{FACTORIZATION}
 within a factor two in the branching ratios.
Then, the hadronic matrix elements of the above operators are given by
\begin{eqnarray}
 \langle K_X\p\mid O_3\mid B\rangle&=&
 \langle K_X\p\mid O_4\mid B\rangle=
{4 \over 3} \langle K_X\p\mid O_5\mid B\rangle=
4\langle K_X\p\mid O_6\mid B\rangle \nonumber\\
&=&{1 \over 3}\langle \p\mid \bar s\gamma_\mu s\mid 0\rangle
\langle K_X\mid \bar s\gamma^\mu (1-\gamma_5) b\mid B\rangle.
\end{eqnarray}
 \noindent
Using the Gordon identity and the color algebra relation,
the operator $O_8'$ reduces to
\begin{equation}
O_8'=i{\alpha_s\over 4 \pi}m_b{1\o q^2}[i{8\o 9} m_b (\bar s_L \r_{\mu}b_L)
(\bar s_L\r^\mu s_L)-{4\o 9}(\bar s\sigma_{\mu\nu}b)
\bar s(p')_L(\r^\mu p'^\nu-\r^\nu p'^\mu)s_L ]
\end{equation}
\noindent
in the limit $m_s=0$, where a Fierz transformation is performed
and the relation $\langle\phi\mid\bar s_Ls_R\mid 0 \rangle=0$ is used.
Then, the hadronic matrix element is factorized as follows:

\begin{eqnarray}
\langle K_X\p\mid O_8'\mid B\rangle&=&i{\alpha_s\over 4\pi}m_b{1\o q^2}\{
i{8\o 9} m_b \langle\p\mid \bar s_L\r_\mu s_L\mid 0\rangle
 \langle K_X\mid \bar s_L \r^{\mu}b_L\mid B\rangle \nonumber\\
&- &{4\o 9}\langle \p \mid \bar s_L(\r^\mu p'^\nu-\r^\nu p'^\mu)s_L
\mid 0\rangle
\langle K_X\mid\bar s\sigma_{\mu\nu}b\mid B\rangle \}\;\; ,
\end{eqnarray}
\noindent
with
\begin{eqnarray}
\langle \p\mid \bar s \r_\mu s\mid 0\rangle &=& g_\p \eta_\mu,  \nonumber\\
 \langle \p\mid \bar s_L(\r^\mu p'^\nu-\r^\nu p'^\mu)s_L\mid 0\rangle&=&
{1\o 4}g_\p(\eta^\mu p_\p^\nu-p_\p^\mu\eta^\nu),
\end{eqnarray}
and where $g_\p=0.23 \G^2$ is taken and $p'^\mu=p_\p^\mu/2$ is assumed.

The hadronic matrix elements
 $\langle K_X\mid\bar s\sigma_{\mu\nu}b\mid B\rangle$
are  given for  the $1 ^{1}S_0$ and  $ 1 ^3S_1$ states
in terms of the form factors by
\begin{eqnarray}
\langle K(p')\mid \bar s\sigma_{\mu\nu}b\mid \bar B(p)\rangle
&=& is^T[(p+p')_\mu (p-p')_\nu-(p-p')_\mu (p+p')_\nu] \ ,\nonumber\\
\langle K^*(p',\epsilon)\mid \bar s\sigma_{\mu\nu}b\mid \bar B(p)\rangle
&=& g_+^T\epsilon_{\mu\nu\lambda\sigma}\epsilon^{*\lambda}(p+p')^\sigma+
 g_-^T\epsilon_{\mu\nu\lambda\sigma}\epsilon^{*\lambda}(p-p')^\sigma
 \\
&+ &h^T\epsilon_{\mu\nu\lambda\sigma}(p+p')^\lambda (p-p')^\sigma
(\epsilon^*\cdot p) \ . \nonumber
\end{eqnarray}
The form factors appearing in the above equations are easily estimated
by  using the relations derived from
heavy quark symmetry
\cite{HQS};
\begin{eqnarray}
 s^T&=&{f_+ - f_-\over 2 m_b} \ ,\nonumber\\
 h^T&=&-{g\over m_b}+{a_+ -a_-\over 2 m_b} \ ,\nonumber\\
 g_+^T - g_-^T  &=& -2 m_b g \ ,\\
  g_+^T + g_-^T &=&{f\over m_b}+2{{p\cdot p'}\over m_b}g, \nonumber
\end{eqnarray}
\noindent
where the form factors $f_+$, $f_-$, $g$, $a_+$ and $a_-$ are defined
in the hadronic matrix elements of the vector and the axial-vector current
\cite{HQS}.
For the $1 ^3P_2$, $2 ^{1}S_0$ and $2 ^3S_1$ states, similar relations are
satisfied,
while for the $1 ^3 P_1$,  $1 ^1P_1$ and $1 ^3P_0$ states,
the overall signs of {\it r.h.s.} in Eq.(19) should be reversed.\par

 These form factors are to be evaluated at $(p-p')^2=m^2_\p$
for each final state and they generally depend on the quark potential.
Here we use the form factors  given by
Isgur, Scora, Grinstein and Wise
\cite{ISGW},
which have been successfully  applied to the electron energy spectra of
the semileptonic $D$ and $B$ meson decays.
These are derived using
harmonic oscillator type wave functions
with a variational method, assuming a Coulomb plus linear
potential.
This simple model gives quite reasonable
 spin-averaged spectra of  $c\bar d$ and $b\bar d$ mesons up to
$L=2$.
These  form factors
 include the relativistic compensation
factor, although their model itself is a nonrelativistic one.
\par
The decay amplitudes are given by

\begin{eqnarray}
\langle  K_X \p \mid H_{eff}\mid B\rangle&=&
{4 G_F\over \sqrt{2}} V_{tb}V_{ts}^*
\{ A \eta^\mu\langle K_X\mid \bar s_L\r_\mu b_L\mid B\rangle \nonumber\\
&+&i B(\eta^\mu p_\p^\nu-p_\p^\mu\eta^\nu)\langle K_X\mid\bar s
\sigma_{\mu\nu}b\mid B\rangle\}\;\; ,
\end{eqnarray}
\noindent with
\begin{eqnarray}
A &\equiv& 6\{C_3(\mu)+C_4(\mu)\}+{3\o 2}\{3C_5(\mu)+C_6(\mu)\}+
{\a_s\o \pi}{m_b^2\o q^2}C_8(\mu), \nonumber\\
 B&\equiv& {\a_s\o 4\pi}{m_b\o q^2}C_8(\mu),
\end{eqnarray}
 \noindent and where $p$ is the three momentum of the $K_X$ meson.

The decay widths of the possible  modes are given in terms of the form factors
by
\begin{equation} \Gamma(B \ra K_X\p)
        ={ G_F^2 g_\p^2 \over 81\pi m_B^2}\mid V_{tb}V_{ts}^*\mid ^2
 p (A^2X_A+B^2 X_B+AB X_{AB})  .
\end{equation}
The quantities $X_A$, $X_B$ and  $X_{AB}$ for each state are given in terms of
the form factors as follows;

\noindent
for the $0^-(1 ^1S_0)$  state,
\begin{equation}
X_A={m_B^2\o m_\p^2} f_+^2 p^2, \quad X_B=16m_B^2m_\p^2 s^{T2} p^2, \
 X_{AB}=-8m_B^2f_+s^T p^2,
\end{equation}
\noindent
for the $1^-(1 ^3S_1)$ state,
\begin{eqnarray}
 X_A&=&{1\o 2} (f^2+4m_B^2g^2p^2) +
   {1\o 4m_{K^*}^2 m_\p^2}\{(p_{K^*}\cdot p_\p)f+   2m_B^2 a_+p^2\}^2
  \ ,   \nonumber\\
 X_B&=&32 m_B^2 g_+^{T2} p^2, \quad  X_{AB}=8m_B^2 g g_+^T p^2,
\end{eqnarray}
\noindent
for the $2^+(1 ^3P_2)$ state,
\begin{eqnarray}
 X_A={m_B^2 \o 4m_{K_2}^2} (k^2&+&4m_B^2h^2p^2)p^2+
 {m_B^2 \o 6m_{K_2}^4 m_\p^2}\{(p_{K_2}\cdot p_\p)k+
  2m_B^2 b_+p^2\}^2 p^2  \ ,\nonumber\\
 & &X_B=16 {m_B^4\o m_{K_2}^2} g_{2+}^{T2} p^4, \quad
    X_{AB}=4{m_B^4\o m_{K_2}^2} h g_{2+}^T p^4,
\end{eqnarray}
\noindent
for the $1^+(1 ^3P_1)$ or the $1^+(1 ^1P_1)$ states,
\begin{eqnarray} X_A&=&{1\o 2} (\ell^2+4m_B^2 q^2p^2)+
{1\o 4m_{K_1}^2 m_\p^2}\{(p_{K_1}\cdot p_\p)\ell + 2m_B^2 c_+p^2\}^2,
\nonumber\\
 X_B&=&8\{m_\p^2(g_{1+}^T-g_{1-}^T)-2(p_{K_1}\cdot p_\p)g_{1+}^T\}^2
  \nonumber\\
&+ & {4\o m_{K_1}^2m_\p^2}[\{m_\p^2(g_{1+}^T-g_{1-}^T)-
2(p_{K_1}\cdot p_\p)g_{+1}^T\}
(p_{K_1}\cdot p_\p)\nonumber\\
&+ &2m_B^2(g_{1+}^T-m_\p^2 h_1^T)p^2]^2 \ ,\nonumber\\
 X_{AB}&=&-{2\o m_{K_1}^2m_\p^2}\{(p_{K_1}\cdot p_\p)\ell
+ 2m_B^2 c_+p^2\} \times    \nonumber\\
& & [\{(g_{1+}^T-g_{1-}^T)m_\p^2-2(p_{K_1}\cdot p_\p)g_{1+}^T\}
  (p_{K_1}\cdot p_\p)+2m_B^2(g_{1+}^T- m_\p^2 h_1^T) p^2] \nonumber\\
&- &  4\ell\{m_\p^2(g_{1+}^T-g_{1-}^T)-2(p_{K_1}\cdot p_\p)g_{1+}^T\},
\end{eqnarray}
\noindent
and for the $0^+(1 ^3P_0)$ state,
\begin{equation} X_A={m_B^2\o m_\p^2} u_+^2 p^2, \quad X_B=0, \ X_{AB}=0.
\end{equation}
For the $0^-(2 ^1S_0)$ and  $1^-(2 ^3S_1)$ states, similar forms
of $X_A$, $X_B$ and $X_{AB}$ as for
 $0^-(1 ^1S_0)$ and  $1^-(1 ^3S_1)$
apply, respectively.
The explicit forms of the form factors
$f_+$, $f$, $g$, $a_+$, $k$, $h$, $b_+$, $\ell$, $q$, $c_+$ and $u_+$
are given in the Appendix of Ref.\cite{ISGW}.
The form factors
$s^T$, $g^T_{+}$, $g^T_{1+}$, $g^T_{2+}$ and $h_1^T$ are calculated
by using Eqs.(19) and the form factors referred to above.
\par
The decay branching ratios are shown
in Table 2 in the SM, where $C_i(\mu)$ are evaluated
at $\mu=m_b=4.58$GeV.
\vskip 0.5cm
\begin{center}
 \unitlength=0.8cm
 \begin{picture}(2.5,2.5)
  \thicklines
  \put(0,0){\framebox(3,1){\bf Table 2}}
 \end{picture}
\end{center}
\vskip 0.5cm
\noindent
It is found that
the decays $B \ra K_1(1400)\p$
 and $B \ra K(1460)\p$ dominate the total decay rate of $B \ra K_X\p$.
We also show the predicted values ``without'' QCD corrections, which cases
were investigated by two of the present authors previously in the second
reference of Refs.
\cite{DAVIES}.
It should be  remarked that the QCD effects
significantly increase the branching ratios
by almost a factor of three.
This is to be compared to the process
$b \ra s\r$, in which the QCD corrections enhance the  decay ratio
by one order as discussed in paper I.
Since the predicted values depend on the values of the form factors,
 we compare our predicted branching ratios for
$K\p$ and $K^*(890)\p$ final states
with the ones calculated using another model of the form factors
given by Bauer, Stech and Wirbel
\cite{BSW}.
As shown in Table 3,
the predicted values
in the latter model are almost three times
larger than the ones in the GSIW model.
Thus, the predicted values depend significantly on the model
of the form factors, and it would obviously be of great value to improve
the reliability of choosing these form factors.
\vskip 0.5cm
\begin{center}
 \unitlength=0.8cm
 \begin{picture}(2.5,2.5)
  \thicklines
  \put(0,0){\framebox(3,1){\bf Table 3}}
 \end{picture}
\end{center}
\vskip 0.5cm
\noindent
These form factors will possibly be tested by the radiative rare decay $B \ra
K^*(890)\r$,
which is the process discussed in the next section.
\par
Finally, we  study the effect of new physics on these decays.
As a typical example of such
new physics, the effect of charged Higgs bosons is investigated.
Then, $C_i(\mu)$ are modified as discussed in section 2.
We show the  predicted branching ratios
for the typical case where $m_H=100$GeV and $\cot\b=1$ in Table 2, where we see
that
the effect of the charged Higgs boson
increases the decay rate at most by $10\sim 18\%$.
This is the extreme case to see the Higgs effect and seems to be
excluded by recent CLEO experiment
\cite{CLEO}
 as discussed in the next section.
Then the charged Higgs boson appears to
have a minor contribution to these decays.
Thus, it may be difficult to observe the evidence of the new physics
in these decays if we take into account the
large ambiguity in the choice of the form factors, as well as the assumption
of factorization.
However, the radiative exclusive rare decays
such as $B \ra K^*(890)\r$ may
help to choose between models of the form factors,
and the factorization assumption will be studied precisely in the normal
 non-leptonic decays of the B meson.
Then the theoretical calculations
of these rare decays will become more reliable, and the contribution of
effects due to new physics  will be able to be discussed quantitatively.
\vskip 1.5cm
\section{The renewed analysis based on the new
CLEO experimental results}
\hspace*{0.6cm}
Recently, the new experimental limit on $B(b \ra s \gamma)$
and the new  value of $B(B \ra K^* \gamma)$ have been reported
by the CLEO group\cite{CLEO} as follows;
\begin{equation}
     B(B \ra X_s\gamma)<5.4 \times 10^{-4},
\end{equation}
\begin{equation}
     B(B \ra K^* \gamma)=(4.5\pm1.5\pm0.9) \times 10^{-5}.
\end{equation}
The experimental data on these processes
improve the previous constraints
\cite{ETC}
on the charged Higgs mass and the parameter $\cot \beta$
in the THDM.
In Fig.1, we show the modified allowed
 parameter region of $\cot \beta$ versus $m_H$ obtained
from the new upper limit
 of $B(B \ra X_s \gamma)$, which replaces that of the Fig.2 of
paper I.
\vskip 0.5cm
\begin{center}
 \unitlength=0.8cm
 \begin{picture}(2.5,2.5)
  \thicklines
  \put(0,0){\framebox(3,1){\bf Fig.1}}
 \end{picture}
\end{center}
\vskip 0.5cm
The value of $m_H$ is restricted to be larger than about 220Gev which
is for $\cot\beta=0,m_t=140$GeV and $m_b=4.58$GeV.
It is of interest to note that
the new experimental upper bound $B(B \ra X_s\gamma)<5.4
\times 10^{-4}$  is very close to the QCD corrected predicted value of
$3.3 \times 10^{-4}$(for $m_t=140$GeV)
in the SM.

The predicted branching fractions of $B \ra K^*(892)\r$
given in the paper I are
\begin{equation}
B(B \ra K^*(892)\r)=\cases{3.0 \times 10^{-5} &  \cr
(5.3\sim 6.9) \times 10^{-5} & (THDM:$m_H=200\G,\cot\beta=0.0\sim 2.0$)}
\end{equation}
at $m_t=140\G$.
These values are consistent with the recent CLEO experimental results given
in Eq.(29).
At present it is hard to distinguish the effect of the charged Higgs boson,
because of large experimental errors.
In our calculation, for both cases, the ratio of the
exclusive decay rate of $B\ra K^*(892)\r$ to the
inclusive decay rate of $B\ra X_s\r$ is 7\%.

Further experimental progress on both the inclusive
and the exclusive processes is eagerly anticipated, to help settle
whether the SM correctly describes these decays, or whether we should
take into account the possibility of physics beyond
the SM.
\section{Summary and conclusion}
\hspace*{0.6cm}
 We have analysed the decays $B \ra K_X\phi$
in the SM, and also studied the effect of inclusion of the charged
Higgs contribution.

First we presented the QCD effects for the effective Hamiltonian relevant
for $b \ra s+s+{\bar s}$, which is mainly induced at the one-loop level
through
the gluonic penguin, where the approximately neglected operators in the
previous analyses of the other rare $B$ decays are properly included.
Then the numerical values of the relevant Wilson coefficient
functions $C_i(m_b)$ in the SM as well as in the THDM are summarized,
where we find that the noteworthy differences among the SM and the
THDMs are seen only for the two coefficients $C_7(m_b)$ and $C_8(m_b)$.

In estimating the hadronic matrix elements of the exclusive decays
$B \ra  K_X\phi$, factorization and a specific model for the form factors are
assumed,   the latter being those given in Ref.
\cite{ISGW}, based on the harmonic oscillator type wave functions
with variational method using the  Coulomb plus linear
potential. As well, these form factors are based on relations derived from
heavy quark symmetry.

We obtained the branching ratios of these processes in the standard
model and then we found that the decays $B \ra K_1(1400)\phi$ and $B \ra
K(1460)\phi$ are the dominant modes of $B \ra K_X\phi$.
We also found that,
as in the weak radiative processes $B \ra
X_s\gamma$, the QCD effect significantly increases the branching
ratios by a factor of about three.
However, the predicted values depend significantly on the model of the form
factors,
which we have shown by comparing the results obtained by using two sets
of form factors, those of
Ref.\cite{ISGW} and Ref.\cite{BSW}.
Accurate measurements of
decays such as $B \ra K^*(890)\gamma$ may be helpful to select the
most reliable model of the form factors.
We also studied the effect of the possible existence of charged
Higgs boson contributions to these decays.
However, according to our calculation, it gives only
a minor contribution, increasing
the decay rates at most by 10-18\%.
Due to the theoretical ambiguities of the prediction, such as its dependence
on the model of form factors and the factorization assumption, such minor
contribution makes the clear observation of evidence for the existence
of the charged Higgs contribution difficult.
So weak radiative decays may be more favourable to search for
charged Higgs effects than
the processes we consider here, $B \ra K_X \phi$.
\vskip 0.5cm
{\it note added}:\quad After most of this work was completed,
we have received the related paper by R. Fleischer\cite{FLEI}.

\begin{center}
{\it Acknowledgement}
\end{center}
\noindent
The authors wish to thank Dr G.C.Joshi for
reading the manuscript.

\newpage
\centerline{\bf Table Captions}
{\bf Table 1}:Coefficients $C_3(m_b)$ to $C_8(m_b)$ of the
 interactions Eqs.(6) in the SM
and THDM ``without'' and ``with'' the QCD effect.
$C_1(m_b)$ and $C_2(m_b)$ are given in Eq.(10).
The mass of top quark is fixed to be 140Gev and $m_b$=4.58GeV.
The THDM(1) and THDM(2) results
correspond to the case $m_H$=220GeV, $\cot \beta=0$ and $m_H=300$GeV,
$\cot \beta=1$, respectively.
\vskip 1.5cm
{\bf Table 2}:Branching ratios in the standard model ``with'' and ``without''
 QCD effect.
 The predictions including the contribution from the charged Higgs bosons
($m_H=100$GeV,$\cot\beta=1$)
``with'' QCD are also shown in the last column.
\vskip 1.5cm
{\bf Table 3}:Predicted branching ratios using the ISGW form factor model
\cite{ISGW} and the BSW form factor model\cite{BSW} in the
SM including QCD corrections.
\label{tab2}
\vskip 1.5cm
\centerline {\bf Figure Captions}
{\bf Fig. 1} The allowed region of parameters $m_H$ and $\cot \beta$ in the
THDM.
The allowed region  is to the right hand side of the corresponding line.
The thick and the thin lines correspond to
the cases $m_b=4.58\G$ and $m_b=5.00\G$,
respectively.
The three cases $m_t =$
 110GeV, 140GeV and 170GeV are shown as indicated in the figure.
\newpage
\begin{center}
\begin{tabular}{| c | c | c | c|} \hline
 Coefficient & ``Without'' QCD & ``With'' QCD(at $m_b$) \\ \hline
             & SM          & SM      \\
             & THDM(1)     & THDM(1)  \\
             & THDM(2)     & THDM(2)  \\ \hline
 $C_3$         & 0.008-0.003i       & 0.015-0.006i  \\
               & 0.008-0.003i       & 0.015-0.006i  \\
               & 0.008-0.003i       & 0.015-0.006i  \\ \hline
 $C_4$         & -0.024+0.009i      & -0.038+0.017i \\
               & -0.024+0.009i      & -0.038+0.017i \\
               & -0.024+0.009i      & -0.038+0.017i \\ \hline
 $C_5$         & 0.008-0.003i       & 0.012-0.006i  \\
               & 0.008-0.003i       & 0.012-0.006i  \\
               & 0.008-0.003i       & 0.012-0.006i  \\ \hline
 $C_6$         & -0.024+0.009i      & -0.043+0.017i \\
               & -0.024+0.009i      & -0.043+0.017i \\
               & -0.024+0.009i      & -0.043+0.017i \\ \hline
 $C_7$         & -0.170      & -0.320 \\
               & -0.304      & -0.423 \\
               & -0.282      & -0.407 \\ \hline
 $C_8$         & -0.089      & -0.157 \\
               & -0.216      & -0.247 \\
               & -0.197      & -0.234 \\ \hline
\end{tabular}
\vskip 1.0cm
\centerline{\bf Table 1}
\end{center}
\newpage
\begin{center}
\begin{tabular}{| l | c | c | c| } \hline
 \ \ \ \ \ \ Process & SM ``with'' QCD  & SM ``without'' QCD
 & Including Charged Higgs \\
 &  & &''with'' QCD \\\hline
$B\ra K\p$ \quad & $0.24\times 10^{-5}$ & $0.08\times 10^{-5}$
& $0.27\times 10^{-5}$  \\
$B\ra K^*(890)\p$ & $0.27\times 10^{-5}$ & $0.10\times 10^{-5}$
& $0.31\times 10^{-5}$  \\
$B\ra K^*_2(1430)\p$ & $0.07\times 10^{-5}$ & $0.03\times 10^{-5}$
& $0.08\times 10^{-5}$ \\
$B\ra K_1(1270)\p$ & $0.57\times 10^{-5}$  & $0.16\times 10^{-5}$
& $0.65\times 10^{-5}$ \\
$B\ra K_1(1400)\p$ & $2.05\times 10^{-5}$  & $0.74\times 10^{-5}$
& $2.41\times 10^{-5}$ \\
$B\ra K_0(1430)\p$ & $0.08\times 10^{-5}$  & $0.03\times 10^{-5}$
& $0.10\times 10^{-5}$ \\
$B\ra K(1460)\p$ & $1.21\times 10^{-5}$  & $0.42\times 10^{-5}$
& $1.38\times 10^{-5}$ \\
$B\ra K^*(1410)\p$ & $0.25\times 10^{-5}$ & $0.09\times 10^{-5}$
& $0.29\times 10^{-5}$  \\ \hline
\end{tabular}
\vskip 1.0cm
\centerline{\bf Table 2}
\end{center}
\noindent
\vskip 1.5cm
\begin{center}
\begin{tabular}{| l | c | c | } \hline
 \ \ \ \ \ Process &\ \ \ ISGW model\ \ \  &\ \ \ BSW model\ \ \  \\ \hline
$B\ra K\p$ \quad & $0.24\times 10^{-5}$ & $0.80\times 10^{-5}$   \\
$B\ra K^*(890)\p$ & $0.27\times 10^{-5}$ & $0.92\times 10^{-5}$  \\ \hline
\end{tabular}
\vskip 1.0cm
\centerline{\bf Table 3}
\end{center}

\begin{thebibliography}{99}
\bibitem{COM}
   The full expressions including $Z^0$- and photonic-penguin interactions are
 given ,for example, in reference\cite{BUCHALLA}.
\bibitem{q2}
   This equation is derived kinematically from
the assumption that $s$-quark in the $\phi$-meson
carries the momentum $p_\phi/2$.
\end{thebibliography}
\end{document}